\documentclass[preprint]{vgtc}               

\ifpdf
  \pdfoutput=1\relax                   
  \usepackage{graphicx}                
  \DeclareGraphicsExtensions{.pdf,.png,.jpg,.jpeg} 
\else
  \usepackage{graphicx}                
  \DeclareGraphicsExtensions{.eps}     
\fi%

\graphicspath{{figures/}{pictures/}{images/}{./}} 

\usepackage{microtype}                 
\PassOptionsToPackage{warn}{textcomp}  
\usepackage{textcomp}                  
\usepackage{mathptmx}                  
\usepackage{times}                     
\usepackage{cite}                      
\usepackage{tabu}                      
\usepackage{booktabs}                  

\onlineid{0}

\vgtccategory{Research}

\vgtcinsertpkg

\preprinttext{Author’s version. Final record to appear in 2023 IEEE International Symposium on Mixed and Augmented Reality Adjunct (ISMAR-Adjunct).}

\title{The Impact of Different Virtual Work Environments on Flow, Performance, User Emotions, and Preferences}

\author{ Alicja Kiluk\thanks{e-mail: alicja.kiluk@uni-bayreuth.de}\\ 
      \scriptsize University of Bayreuth %
\and Viktorija Paneva\thanks{e-mail: vpaneva@acm.org}\\ %
        \scriptsize University of Bayreuth   
\and Sofia Seinfeld\thanks{e-mail: sseinfeld@uoc.edu}\\ %
     \scriptsize Universitat Oberta de Catalunya %
\and J{\"o}rg M{\"u}ller\thanks{e-mail: joerg.mueller@uni-bayreuth.de}\\ %
    \parbox{1.4in}{\scriptsize \centering University of Bayreuth \\ 
    }}

\teaser{
  \centering
  \includegraphics[width=\linewidth]{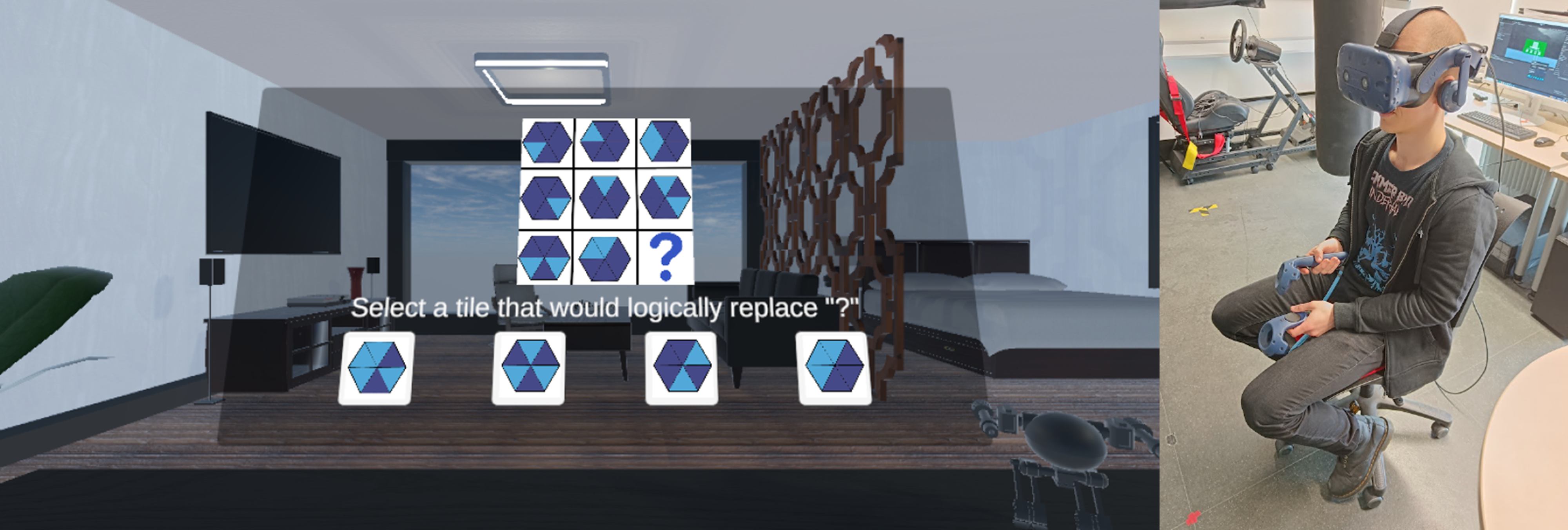}
  \caption{Our study examines how different work environment in VR affect users' flow, performance, emotional state, and preferences through pre-study interviews and a user study with three distinct VR work environments: the Dark Room, the Empty Room, and the Furnished Room. The figure shows a user solving a logic-based multiple-choice question in the Furnished Room environment.}
  \label{fig:teaser}
}

\abstract{
This research explores how different virtual work environments, differing in the type and amount of elements they include, impact users’ flow, performance, emotional state, and preferences. 
Pre-study interviews were conducted to inform the design of three VR work environments:  the Dark Room, the Empty Room, and the Furnished Room.
Fifteen participants took part in a user study where they engaged in a logic-based task simulating deep work while experiencing each environment.
The findings suggest that while objective performance measures did not differ significantly, subjective experiences and perceptions varied across the environments.
Participants reported feeling less distracted and more focused in the Dark Room and the Empty Room compared to the Furnished Room. 
The Empty Room was associated with the highest levels of relaxation and calmness, while the Furnished Room was perceived as visually appealing yet more distracting. 
These findings highlight the variability of user preferences and emphasise the importance of considering user comfort and well-being in the design of virtual work environments. 
The study contributes to the better understanding of virtual workspaces and provides insights for designing environments that promote flow, productivity, and user well-being.
}

\CCScatlist{
  \CCScatTwelve{Human-centered computing}{Human computer interaction (HCI)}{Interaction paradigms}{Virtual reality};
  \CCScatTwelve{Human-centered computing}{Human computer interaction (HCI)}{HCI design and evaluation methods}{User studies}
}

\begin{document}

\firstsection{Introduction}
\maketitle

In recent years, virtual reality (VR) technology has gained significant attention as a tool for creating immersive and interactive experiences in various domains. 
One promising application of VR is the development of virtual workspaces that can simulate real-world offices or study environments. 
These virtual workspaces have the potential to provide users with an enhanced and customisable work experience, free from the limitations and distractions often encountered in physical work environments~\cite{Grubert2018}. 

Flow, characterised by deep concentration and optimal engagement in a task, is considered essential for achieving high performance and a satisfying work experience~\cite{Csikszentmihalyi1990}.
Creating an environment that is free from distractions is crucial for facilitating flow~\cite{Schaffer2013}.
Auffegger et al.~\cite{Aufegger2022} introduced a VR productivity framework that encompasses physical, environmental, cognitive, and behavioural needs, to ensure productivity and organisational growth.
Their study highlights the importance of an immersive and distraction-free environment for asynchronous work tasks, such as programming or writing.

To address distractions in open-plan offices, researchers have investigated the use of augmented reality and visual separators~\cite{Lee2019}.
Ruvimova et al.~\cite{Ruvimova2020} explored the use of VR to mitigate distractions for knowledge workers. 
Participants performed a visual programming task in physical (open and closed plan offices) and VR (beach and virtual office) environments.
While the physical closed office was the most preferred and scored the highest in fostering flow, VR outperformed the physical environment when comparing open-plan offices.
Additionally, guidelines have been proposed to address distractions in virtual experiential learning environments~\cite{Bian2022}.
Hasenbein et al.~\cite{Hasenbein2022} examined how different social-related VR classroom configurations affected the visual attention of six graders towards the presented study material.
A study by Latini et al.~\cite{Latini2021} investigated the effects of wall colour and room temperature on productivity and comfort, in both a real and virtual office setting.

Personalisation has also been a focus in the development of VR workspaces. 
Different personalised learning environments have been explored for memory training across different age groups~\cite{Schmuecker2021}. 
Furthermore, personalised VR content that adapts according to the user's emotional state has been developed for relaxation purposes~\cite{Heyse2020}.

Moreover, VR has shown potential for supporting work in confined spaces and transit environments. 
Li et al.~\cite{Li2022} examined the use of VR for car passengers performing work-related tasks. 
The layout of virtual displays for productivity in shared transit settings, such as trains and planes, has also been studied~\cite{Medeiros2022}.

Despite the growing interest in virtual workspaces and their potential benefits, there is still a significant research gap in our understanding of how different elements within VR environments impact users' flow, performance, and emotions. 
While some studies have investigated the effect of the VR environment on productivity and performance, they predominantly focused on comparing real-world and VR settings.
Only few explored how different types of virtual work environments affect users' performance and emotional states. 
Our study addresses this research gap by systematically investigating the impact of three distinct VR work environments, each presenting varying levels of visual and auditory stimuli. 
We hypothesise that higher flow and performance can be achieved in the environments with less visual and auditory stimuli, as they would allow users to allocate more attention to the task at hand.
By gaining a better understanding of how VR environments affect users' productivity and well-being, we can contribute to the design of virtual workspaces that promote flow, enhance user well-being, and optimise overall work performance.
\begin{figure}[b]
    \centering
    \includegraphics[width=\columnwidth]{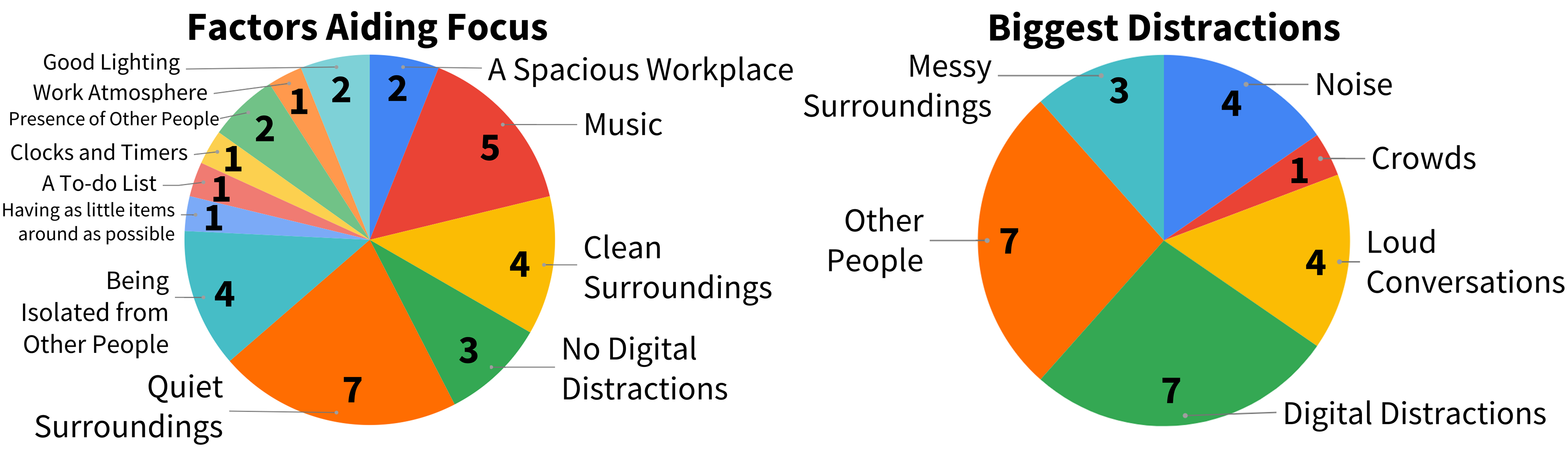}
    \caption{Results of the interviews highlighting the key aspects of the work environment that aid users in maintaining focus (left) and the biggest distractions they encounter (right). The pie chart shows the number of participants that mentioned a given factor in relation to what aids their focus or distracts them when performing deep work.}
    \label{fig:interview_pie_charts}
\end{figure}

\section{Exploring User Perspectives to Inform Virtual Workspace Design}

To gather information about common real-world practices, distractions, and preferred work environments, we conducted some exploratory interviews with users.
The data collected in the interviews informed the design of the virtual workspace environments for our subsequent user study. 
Fifteen people participated in the interviews, aged 18 to 50 (mean 27).
Some of the interviews were conducted in person and some remotely. 
The interview consisted of eight questions. 
The questions are provided in Appendix~\ref{app_interview_questions}.

The participants were asked about workplace factors that help them focus and be productive. A breakdown of their responses is shown in Figure~\ref{fig:interview_pie_charts}. 
The most frequently mentioned aspect was quiet surroundings, with participants highlighting noise as a major distraction and emphasising the importance of silence for staying focused. 
Additionally, five individuals mentioned that listening to music assists their concentration, although preferences varied widely, encompassing genres such as classical, instrumental, electronic, and hip-hop.
Hence no consensus on the type of music preferred for working was observed.
Other mentioned focus-enhancing factors included: isolation from others, clean and distraction-free surroundings, and ample space.

Participants were also asked to identify the most common distractions they encounter while working. 
Figure~\ref{fig:interview_pie_charts} presents a pie chart illustrating the breakdown of their responses. 
The top distractors mentioned were digital distractions and the presence of other people. 
Participants noted that the presence of others makes it challenging to enter a flow state, and the actions of others can be highly distracting. 
Regarding digital distractions, participants highlighted the prevalence of smartphones and laptops in their surroundings, which often lead to distractions like scrolling through social media or browsing the web, both negatively affecting productivity. 
Additionally, participants identified environmental noise as a significant distraction that disrupts their workflow, encompassing factors such as traffic, construction noise, and people speaking loudly.

Overall, participants expressed a preference for a modern, minimalistic office or a natural environment as their ideal VR work environment. Factors such as large windows, good lighting, and a pleasant view were important. 
There was variability in music preferences, with some participants indicating a desire for background music, while others favoured a silent workspace.
Interestingly, participants specifically mentioned not wanting to work from a beach due to its association with relaxation rather than productivity.

The design considerations derived from the interview findings involved the creation of quiet, isolated, and spacious environments to facilitate focused work. 

\begin{figure}[t]
    \centering
    \includegraphics[width=\columnwidth]{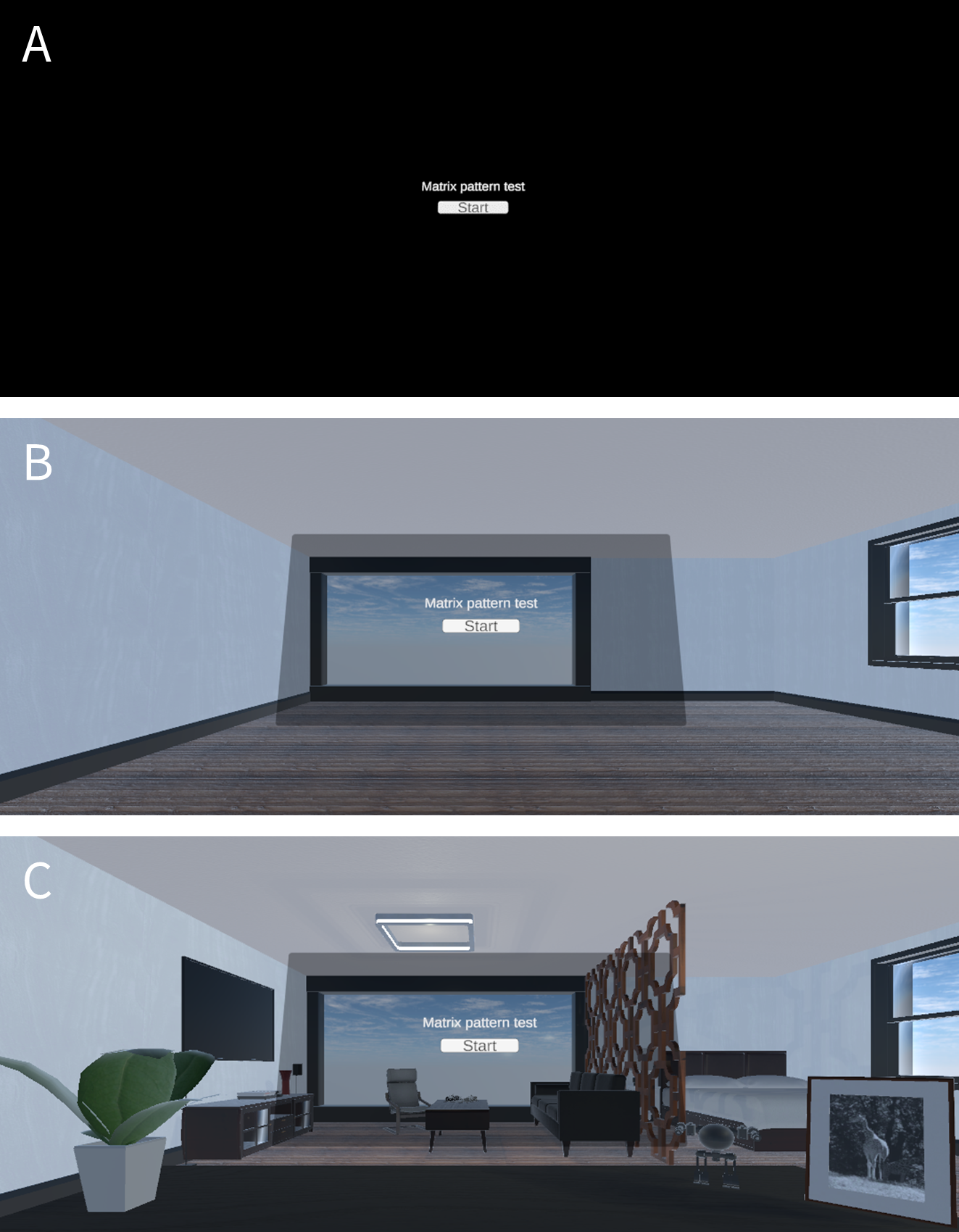}
    \caption{The three different virtual work environments, created for the user study: (A) a distraction-free dark room, (B) an empty room with windows looking to the sky, and (C) a fully furnished modern apartment.}
    \label{fig:conditions}
\end{figure}

\section{The System}

The application for the experiment was created in the Unity game engine, and experienced via an HTC Vive Pro HMD.
Users interacted with the application using hand-held HTC Vive controllers, employing raycasting to select targets on the task screen. 
In the condition with audio, the built-in HTC Vive headphones were used. 
The development and experimentation machines were equipped with a Nvidia GTX1070 graphics card and Ryzen 55600X CPU.

Three different environments were created for the application, depicted in Figure~\ref{fig:conditions}. 
The first environment was a dark room designed to eliminate distractions and ambient sounds.
The second environment was an empty room with a modern design, featuring windows that allowed participants to glimpse outside. 
The skybox was modified to display a bright sky.
The third environment was a fully furnished room, resembling a modern apartment.
It featured a desk with a small potted plant and a photo frame, a sleeping area, and a living area with some other objects and decorations.
Gentle sounds of wind blowing outside were added to the environment to provide some natural ambience. 

\section{User Study}

\subsection{Experimental Design}
The user study consisted of a within-groups experimental design. 
The participants experienced three possible types of VR work environments: 1) Dark Room, 2) Empty Room, and 3) Furnished Room. 
The environments were presented in a randomised order.
The experiment lasted an approximate of 1 hour per participant.

\subsection{Participants}
Fifteen participants (4 female, 11 male) took part in the study, aged between 19 and 33 (mean 25.87, SD 3.44). 
Before the experiment started, the participants read an
information sheet about the study and signed a consent form.
The study was approved by the Ethical Committee of the University of Bayreuth and followed ethical standards as per the Helsinki Declaration. 
All participants received a monetary reimbursement for their participation.

\subsection{Experimental Task}
In selecting the task for the experiment, we aimed to simulate the type of deep work typically encountered in office or study settings. 
Our goal was to create an activity that required participants to engage in focused thinking and problem-solving, reflecting the cognitive demands of professional or academic work. 
It was crucial to find a task that struck the right balance - challenging enough to promote cognitive engagement, but without demanding constant visual attention and user reaction, such that the environment becomes irrelevant.

To meet these criteria, we chose logic-based multiple-choice questions, similar to~\cite{Raven2003, Hill2005}. 
Each question featured a matrix of symbols, where the bottom right cell was marked with a question mark. 
Participants had to identify the symbol that should replace the question mark to complete the pattern within the matrix. 
Figure~\ref{fig:teaser} illustrates a sample question used in the experiment for training purposes.
Participants were given a mix of 2x2 and 3x3 matrix pattern questions, with four possible answers provided below each question.

During the task, if participants selected an incorrect answer, a red pop-up appeared, while a green pop-up indicated a correct answer. 
The application comprised a total of 60 questions, randomly divided into three subsets of 20 questions for each condition. 
For each correctly solved question, participants earned 1 point. Hence the maximum possible score in each environment was 20 points.
After completing the task in an environment, an end screen was displayed, indicating to participants that it was time to take a break.

\subsection{Measures}
To assess participants' performance, we measured completion time and task scores in relation to the logic-based multiple-choice questions.
Qualitative data was gathered through questionnaires that captured participants' emotional state and work flow in relation to each environment. 
The Self-Assessment Manikin (SAM)~\cite{Bradley1994} and Flow Short Scale (FSS)~\cite{Jackson2004} questionnaires were used for this purpose.
SAM utilised graphical representations to measure pleasure, arousal, and dominance.
The FSS consisted of 13 questions answered on a 7-point Likert scale. 
Additionally, four study-specific questions, structured similarly to the FSS were administered.
Lastly, the study was concluded with a semi-structured interview, where participants had the opportunity to talk about their experience in each environment and express preferences.

\subsection{Procedure}
Upon arrival to the university lab, participants were greeted and provided with information about the study. 
They were encouraged to ask any questions or seek clarifications regarding the procedure. Once all their inquiries were addressed, the participants were asked to sign a consent form and complete a short demographics questionnaire.
Following this, the experimenter guided them to a designated area and helped them adjust the VR headset to their individual preferences, ensuring their comfort. 
The subsequent training session allowed the participants to familiarise themselves with the virtual surroundings and the experimental task. 
They were presented with an example matrix pattern question to practice. 
Once they successfully completed the training session, they progressed to the main part of the experiment.
The main study consisted of three sessions, one for each experimental condition. 
In each session, participants were assigned the task of solving a series of 20 matrix pattern questions. 
After completing each session, they were instructed to take off the HMD and take a break. 
During these breaks, participants were asked to fill out the SAM and FSS questionnaires, as well as provide answers to the additional study-specific questions.
Upon completing all three sessions, a semi-structured interview was conducted.

\subsection{Results}

\subsubsection{Task Scores}

\begin{figure}[h]
    \centering
    \includegraphics[width=0.75\columnwidth]{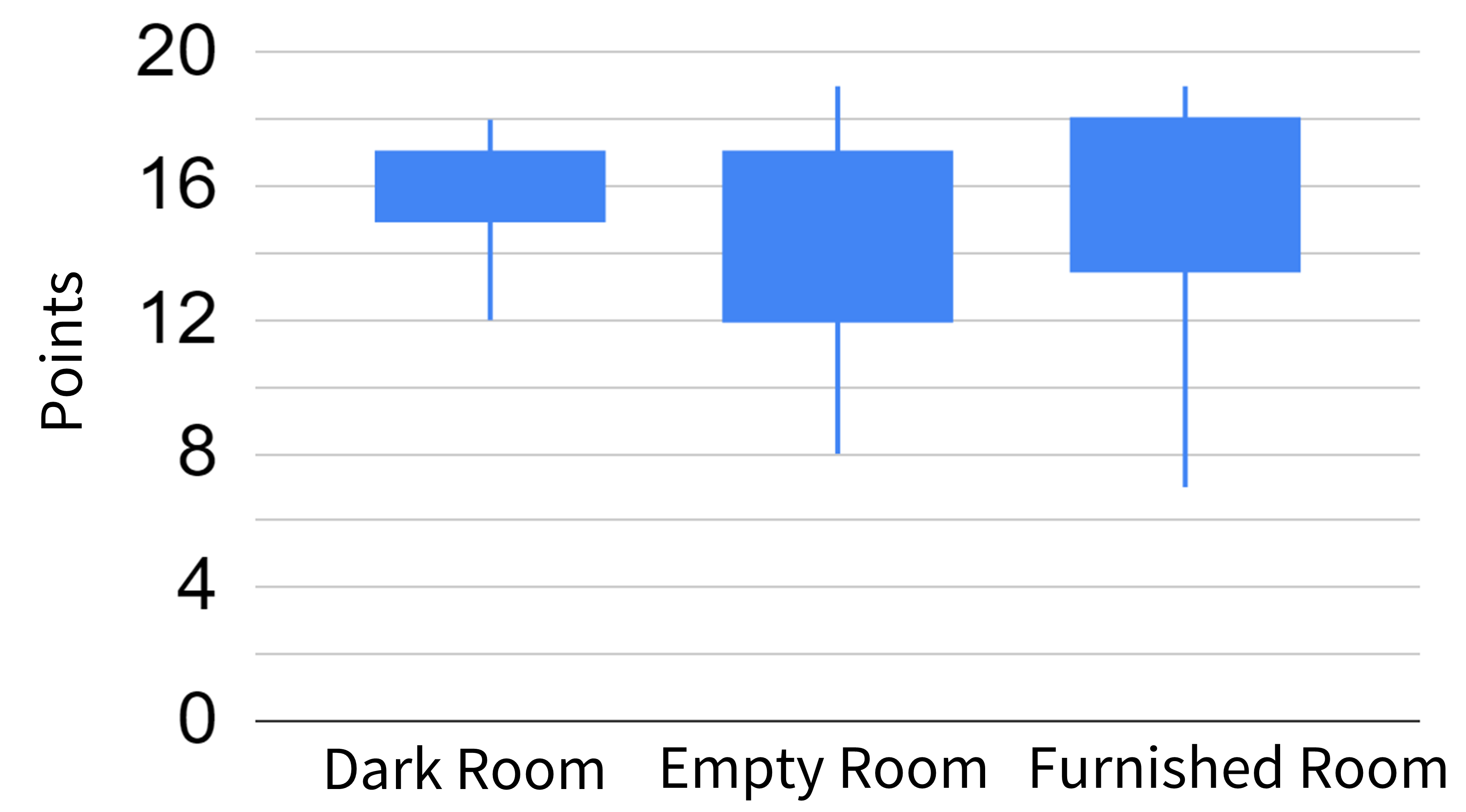}
    \caption{Boxplot of the total task scores for each virtual work environment: Dark Room, Empty Room, and Furnished Room. }
    \label{fig:task_score}
\end{figure}

The average score was 15.9 (SD 1.9) for the Dark Room, 14.8 (SD 3.4) for the Empty Room, and 14.9 (SD 3.6) for the Furnished Room. 
The results are shown in Figure~\ref{fig:task_score}. 
Using the Friedman test, no significant difference in the scores between the virtual work environments was found ($\chi^2=0.94$, $df=2$,  $p=0.62$), despite the Empty and Furnished Room having larger standard deviations than the Dark Room.

\begin{figure}[h]
    \centering
    \includegraphics[width=0.75\columnwidth]{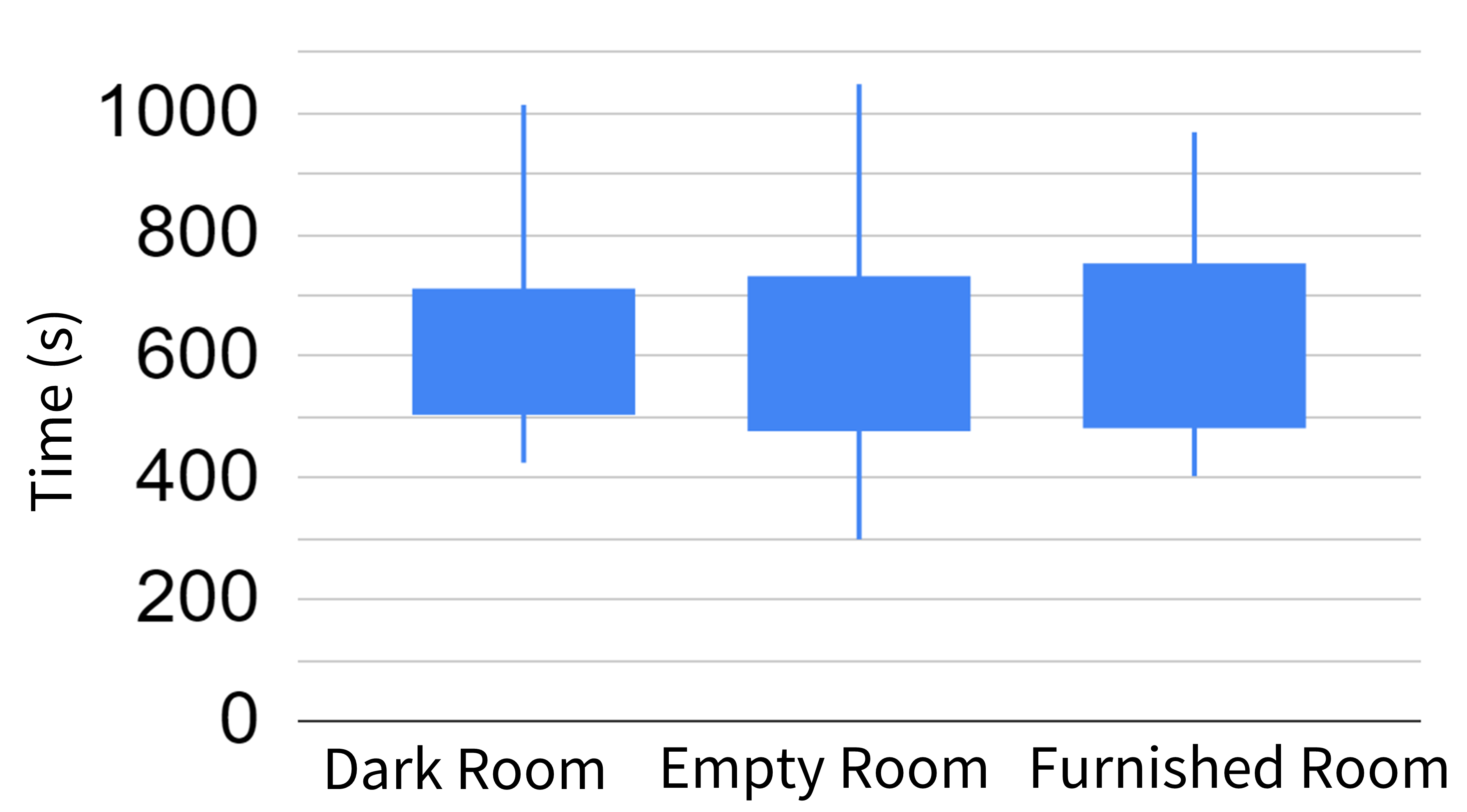}
    \caption{Boxplot of the total task completion time for each virtual work environment: Dark Room, Empty Room, and Furnished Room. }
    \label{fig:task_time}
\end{figure}

\subsubsection{Completion Time}

The average task completion time was 642.6s (SD 191.2) for the Dark Room, 631.6s (SD 190.0) for the Empty Room, and 643.6 (SD 191.4) for the Furnished Room. 
The results are shown in Figure~\ref{fig:task_time}. 
The ANOVA test indicated no significant differences
between the three (p=0.96). 


\subsubsection{SAM}

\begin{figure}[b]
    \centering
    \includegraphics[width=\columnwidth]{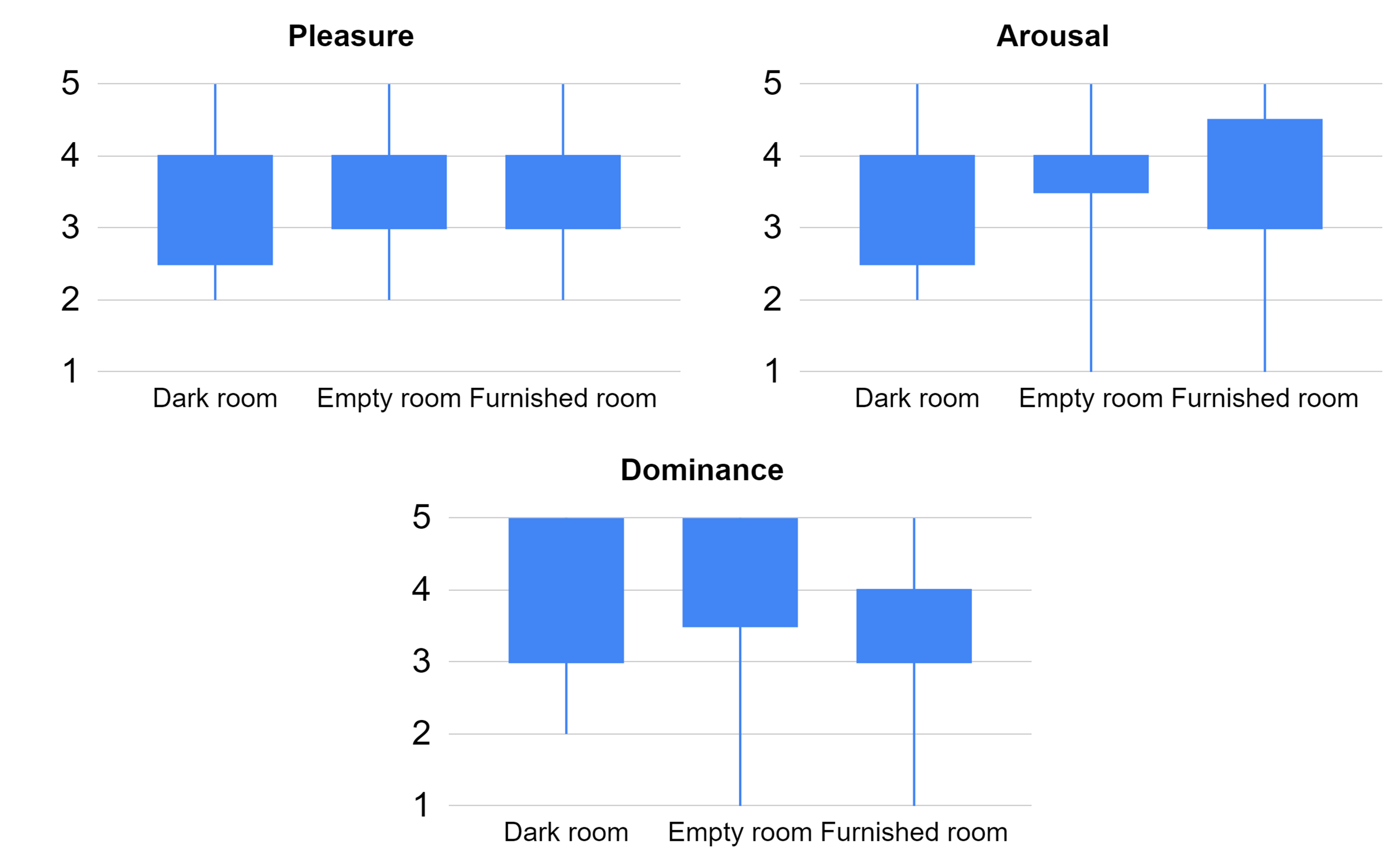}
    \caption{Boxplot of the three dimensions of the SAM questionnaire (pleasure, arousal, and dominance) for each of the three tested VR work environments: Dark Room, Empty Room, and Furnished Room.}
    \label{fig:SAM}
\end{figure}

For the analysis, the participants' responses to the SAM questionnaire were mapped to a 5-point Likert scale, with 1 representing the lowest score answer and 5 representing the highest score answer.
The responses for each dimension: pleasure, arousal, and dominance and in each VR work environment are illustrated in Figure~\ref{fig:SAM}. 
Overall, the participants' answers were relatively similar across the environments. 
The average pleasure score was 3.26 (SD 0.96) for the Dark Room, 3.60 (SD 0.83) for the Empty Room, and 3.53 (SD 0.99) for the Furnished Room. 
The average arousal score was 3.40 (SD 1.06) for the Dark Room, 3.67 (SD 0.90) for the Empty Room, and 3.53 (SD 1.24) for the Furnished Room. 
Finally, the average dominance score was 3.93 (SD 1.10) for the Dark Room, 3.87 (SD 1.19) for the Empty Room, and 3.47 (SD 1.12) for the Furnished Room. 
The Friedman test did not indicate any significant differences in the pleasure ($\chi^2=1.86$, $df=2$,  $p=0.40$), arousal ($\chi^2=0.22$, $df=2$,  $p=0.89$), and dominance ratings ($\chi^2=3.04$, $df=2$,  $p=0.21$) between the environments.

\subsubsection{Flow Short Scale}

\begin{figure}[h]
    \centering
    \includegraphics[width=\columnwidth]{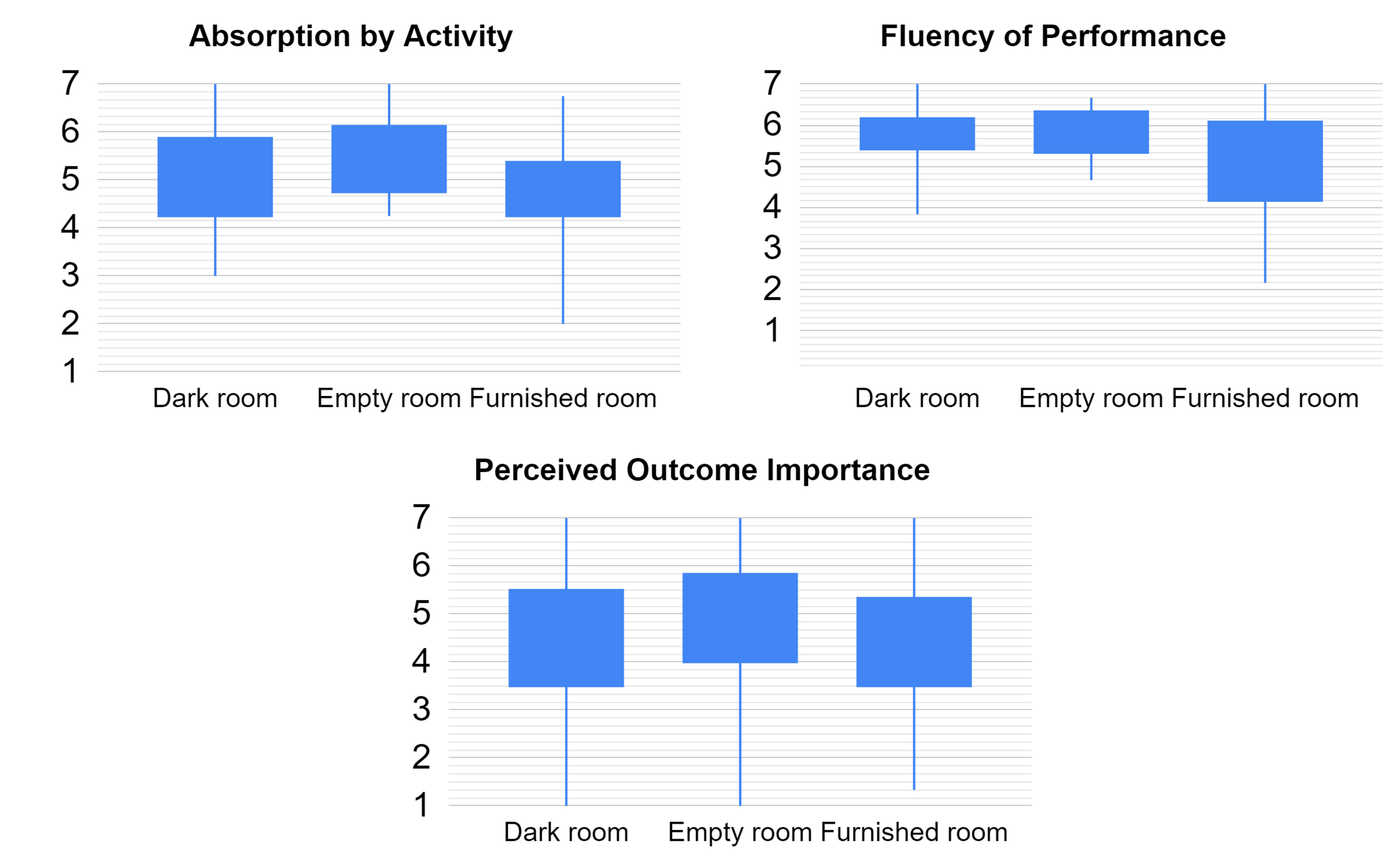}
    \caption{Boxplot of the three dimensions of the FSS questionnaire (absorption by activity, fluency of performance, perceived outcome importance) for each virtual work environment: Dark Room, Empty Room, and Furnished Room.}
    \label{fig:FSS}
\end{figure}

The FSS questionnaire consisted of 13 statements which participants rated on a scale from 1 (not at all) to 7 (very much). 
The questionnaires measures flow based on the following three dimensions: absorption by activity, fluency of performance, and perceived outcome importance.
The results per dimension and condition are shown in Figure~\ref{fig:FSS}.
The average score for \textit{absorption by activity} was 5.13 (SD 1.15) for the Dark Room, 5.40 (SD 0.86) for the Empty Room, and 4.80 (SD 1.15) for the Furnished Room. 
The average score for \textit{fluency of performance} was 5.68 (SD 0.85) for the Dark Room, 5.79 (SD 0.67) for the Empty Room, and 5.03 (SD 1.34) for the Furnished Room. 
Lastly, the average score for \textit{perceived outcome importance} was 4.49 (SD 1.75) for the Dark Room, 4.71 (SD 1.67) for the Empty Room, and 4.49 (SD 1.59) for the Furnished Room. 
The Friedman test did not indicate any significant differences  between the environments for \textit{absorption by activity} ($\chi^2=4.51$, $df=2$,  $p=0.10$), \textit{fluency of performance} ($\chi^2=4.03$, $df=2$,  $p=0.13$), and \textit{perceived outcome importance} ($\chi^2=0.59$, $df=2$,  $p=0.74$).

\subsubsection{Additional Questions}

\begin{figure}[t]
    \centering
    \includegraphics[width=\columnwidth]{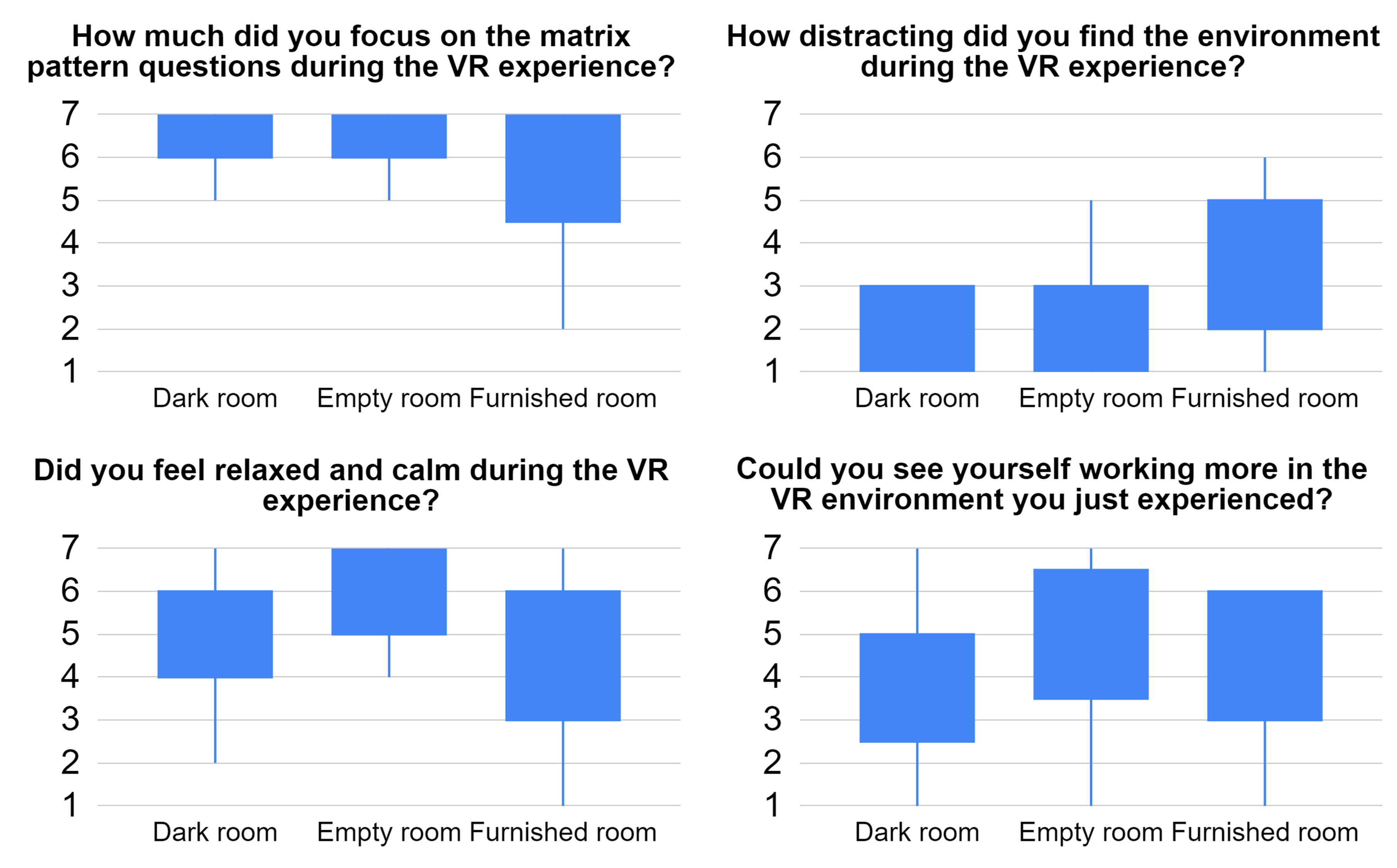}
    \caption{Boxplot of the answers to the four additional study-specific questions for each virtual work environment: Dark Room, Empty Room, and Furnished Room.}
    \label{fig:4_questions}
\end{figure}

After the SAM and FSS questionnaires, we administered four additional questions specific to our study. 
These questions were also answered on a scale from 1 (not at all) to 7 (very much). 
Box plots representing the answers to those questions are presented in Figure~\ref{fig:4_questions}. 
The average score for the first question \textit{How much did you focus on the matrix patter questions during the VR experience?} was 6.47 (SD 0.74) both for the Dark Room and the Empty Room, and 5.47 (SD 1.64) for the Furnished Room. 
Using the Friedman test, we observed a significant difference between environments ($\chi^2=7.70$, $df=2$,  $p=0.021$).
A post-hoc test using Wilcoxon tests with false discovery rate correction showed the significant differences between the Furnished Room and the Dark Room (p=0.035), and between the Furnished Room and the Empty Room (p=0.035).
The average score for the question \textit{How distracting did you find the environment during the VR experience?} was 1.80 (SD 0.94) the Dark Room, 2.00 (SD 1.31) the Empty Room, and 3.40 (SD 1.72) for the Furnished Room. 
A Friedmann test indicated that there was a significant difference between conditions ($\chi^2=10.17$, $df=2$,  $p=0.0062$), showing that the Furnished Room was considered more distracting compared to the rest of conditions. 
However, such difference did not survive the post-hoc pairwise comparison (p=0.06).
In the third question, we asked participants how relaxed and calm they felt during each condition, and obtained an average score of 4.73 (SD 1.62) the Dark Room, 5.80 (SD 1.15) the Empty Room, and 4.53 (SD 1.85) for the Furnished Room. 
The Friedmann test indicated a significant main effect of the virtual environment ($\chi^2=10.11$, $df=2$,  $p=0.0064$).
The Wilcoxon tests with false discovery rate correction showed the significant differences between the Empty and the Furnished Room (p=0.0067), and also indicated a possible trend between the Empty and the Dark Room (p=0.051).
In the last question, we asked participants whether they could see themselves working in the given environment in the future.
The average score for the Dark Room was 3.93 (SD 1.94), 5.07 (SD 2.12) for the Empty, and 4.20 (SD 1.74) for the Furnished Room.
The Friedman test indicated no significant differences between the three ($\chi^2=5.59$, $df=2$,  $p=0.061$).
However, we observe a trend suggesting a potential pattern or tendency among the conditions. 
To confirm the significance of this trend, further investigation with a larger sample size is necessary.

\subsubsection{Semi-structured Interview}

The final part of the data collection process involved conducting semi-structured interviews with each participant at the end of the user study. 
The interview questions can be found in Appendix~\ref{app_end_interview}. 

The first question of the interview asked for participants' opinions on each of the environments they experienced. 
They were encouraged to give their thoughts about them and describe things they liked and disliked. 
Their responses were categorised into positive, neutral, and negative based on the predominant themes.
The Dark Room had an almost even split of positive and negative answers with 6 and 7, respectively. Only 2 of the opinions were marked as neutral. 
This shows that participants either liked the dark environment or really disliked it, which is backed up by their comments. 
Those who liked the Dark Room stated that they felt like nothing could distract them and they could fully immerse themselves in the task. 
They thought it was a comfortable and good place to work in. 
Some also commented that they enjoyed the large contrast between their environment and the task. 
On the other hand, people who did not enjoy the dark environment described it as stressful and oppressive.
4 people stated that they felt stressed and worried while in this environment, with 1 person expecting a jump-scare to appear at some point. 
A few people also mentioned that there was a stark contrast between the environment and the task, making it harder to focus and inducing feelings of fatigue.

Opinions on the Empty Room were predominantly positive, with 7 people having a positive and 5 having a neutral opinion about it. 
Those who liked the Empty Room said that the bright and spacious environment made them feel comfortable, while the lack of objects around them helped them concentrate. 
Only 3 people said that they disliked it. 
These participants mentioned that the room seemed unfinished and lacked objects.
They reported that they felt uncomfortable sitting in a space without any objects and commented on the room feeling disproportionately large while they felt small.

Lastly, the Furnished Room received the most neutral answers (7), as well as 5 positive opinions, and 3 negative. 
Most people said that they really liked the environment and it evoked the feelings of being in a modern, clean apartment. 
6 people were bothered by the sounds of the wind, noting them as a major distraction, while 1 person said that it helped them feel relaxed during the experiment. 
Many people also found that the large amount of objects surrounding them in the VR environment caused them to get distracted quite often.
Interestingly, 2 participants stated that despite the Furnished Room not being a place where it was the easiest to focus, they would still choose it over the other two environments as it made them feel more comfortable. 
They elaborated that, in the long run, they would be willing to trade a minor potential improvement in performance in order to work in an environment that offered both comfort and aesthetic appeal.

Participants were also asked to identify the environments where they found it easiest and hardest to focus. 
The responses to the question \textit{In which of the environments did you find it the easiest to focus?} were evenly divided, with 7 participants choosing the Empty Room and 7 choosing the Dark Room. Only 1 participant felt that the Furnished Room was the most conducive to focus. 
In terms of the worst environment for getting into the flow, 8 participants selected the Furnished Room. 
This finding aligns with the comments highlighting the distractions posed by the objects and the sounds of the wind in that environment. On the other hand, 3 participants found the Empty Room to be the most challenging for focus, while 4 participants chose the Dark Room. 

In one of the final questions of the interview, participants were asked to select their preferred work environment out of the three options. 
Out of the 15 participants, 8 chose the Empty Room, appreciating its bright and comfortable atmosphere. 
Four participants favoured the Furnished Room, considering it the best choice for long-term work. 
Only 3 participants expressed a preference for the Dark Room. 
However, some participants noted that their choice of work environment could vary depending on their mood or the type of task they were assigned.

Lastly, participants had the opportunity to provide additional feedback and comments about the experiment, which will be discussed in the next section.

\section{Discussion}


The results of the statistical analysis did not reveal any significant differences between the three virtual environments in terms of quantitative performance metrics, as well as SAM and FSS questionnaire scores. 
However, participants' responses to study-specific questions indicated that they perceived their focus to be higher in the Dark and the Empty Room compared to the Furnished Room,  while also perceiving the Furnished Room as the most distracting.
Interestingly, participants reported feeling the most relaxed and calm in the Empty Room.

Despite assigning it a lower rating than the other environments, many participants expressed a preference for the Furnished Room. 
They found it visually appealing, comfortable, and a place they would enjoy living in. 
Two participants even indicated a willingness to prioritise comfort and aesthetics over optimal performance. 
These findings emphasise the importance of considering user well-being when designing virtual workspaces. 
This is in line with recent research indicating that incorporating elements in the virtual environment that promote well-being, specifically virtual plants, positively influenced performance in short-term memory and creativity tasks~\cite{Mostajeran2023}.

Participants also noted that their preferred work environment might vary depending on the nature of the task. 
For simpler tasks without tight deadlines, participants preferred comfortable and relaxing spaces where they could work at their own pace.
Conversely, for more challenging and demanding tasks, environments like the Empty or the Dark Room were deemed more suitable for enhancing focus and immersion, aligning with existing literature~\cite{Grubert2018, Schaffer2013}. 

While our initial hypothesis, positing that environments with fewer visual and auditory stimuli would foster better performance and flow, was not confirmed, we note that numerous different factors affecting individuals in different ways could have contributed to this outcome. 
The interplay of many factors in achieving the state of flow complicates the isolation of individual effects~\cite{Schaffer2013}. 
Further investigation is needed, particularly concerning the impact of personal preferences on performance. 
For instance, previous research has indicated that the limited ability to personalise one’s physical work area can have a negative effect on occupant perceived productivity and health~\cite{Kim2016}. 
Furthermore, it is plausible that the selected experimental task might lack the necessary sensitivity to detect such effects. 
In our study, we employed logic-based matrix completion tasks as proxies for deep work tasks encountered in professional and educational settings. It is possible, however, that results might differ for different task types, levels of difficulty, and prolonged session durations, warranting further investigation.
To guide future research, bellow we outline some possible improvements and identify promising directions:

1. Future studies could explore a range of cognitive tasks, including some that require more creative skills. 
Additionally, incorporating tasks with varying levels of difficulty would provide insights into how the environments influence performance across different levels of task complexity. 
This would provide insights into the optimal design of virtual workspaces to support individuals' cognitive abilities and performance across varying task demands.

2. Increasing the sample size is necessary to allow for clearer conclusions regarding significant differences in flow, performance, and preferences between the different VR work environments.

3. Accurate eye-tracking could provide additional useful information.
The data we collected from the headset tracking feature of the application was often unreliable due to inconsistencies between headset orientation and participants' line of sight.
This data had to be excluded from the analysis.
Implementing eye-tracking methods instead would provide more accurate data, eliminating, for example, the impact of participants wearing the HMD slightly tilted to the side.
Moreover, it would allow to measure and differentiate the time participants spend focusing on the task at hand from the time they spend observing their surroundings.

4. Participants suggested providing additional in-task progress feedback, such as displaying the correct answer when a question was incorrectly answered, and adding a progress bar that indicates the number of remaining questions in the particular session.
Furthermore, offering customisation options was requested by some participants. 
Some would have liked the option to move and resize the task window for better visibility. Others wished to have the option to change colours and contrast within the task window.


\section{Conclusion}
This research investigated the impact of different virtual work environments on users' flow, performance, emotional state, and preferences. 
The user study included three different virtual environments: the Furnished Room, the Dark Room, and the Empty Room.
While no significant differences were found in objective performance measures, as well as pleasure, arousal, and dominance between the three virtual environments, subjective experiences and perceptions varied among participants.
Participants reported feeling less distracted and more focused on the task in the Empty and Dark Rooms compared to the Furnished Room. 
Moreover, they felt the most relaxed and calm in the Empty Room. 
Further research is recommended to validate and expand upon these findings. 
Increasing the sample size, implementing accurate eye-tracking methods, exploring a wider variety of tasks with varying levels of difficulty, and providing additional in-task feedback and customisation options are suggested for future studies. 
The findings from this study contribute to the growing body of knowledge on VR workspaces and provide practical implications for their design. 
By understanding the factors that facilitate focus, identifying common distractions, and considering user preferences for work environments, designers can create virtual workspaces that promote productivity and enhance user well-being.

\acknowledgments{The authors would like to thank all participants in the study.}

\bibliographystyle{abbrv-doi}
\bibliography{template}

\begin{thebibliography}{10}

\bibitem{Aufegger2022}
L.~Aufegger and N.~Elliott-Deflo.
\newblock Virtual reality and productivity in knowledge workers.
\newblock {\em Frontiers in Virtual Reality}, 3, 2022.

\bibitem{Bian2022}
Y.~Bian, C.~Zhou, J.~Liu, W.~Geng, and Y.~Shi.
\newblock The effect of reducing distraction on the flow-performance link in
  virtual experiential learning environment.
\newblock {\em Virtual Reality}, 26(4):1277--1290, 2022.

\bibitem{Bradley1994}
M.~M. Bradley and P.~J. Lang.
\newblock Measuring emotion: The self-assessment manikin and the semantic
  differential.
\newblock {\em Journal of Behavior Therapy and Experimental Psychiatry},
  25(1):49--59, 1994.

\bibitem{Csikszentmihalyi1990}
M.~Csikszentmihalyi.
\newblock Flow. the psychology of optimal experience.
\newblock {\em New York (HarperPerennial)}, 1990.

\bibitem{Grubert2018}
J.~Grubert, E.~Ofek, M.~Pahud, and P.~O. Kristensson.
\newblock The office of the future: Virtual, portable, and global.
\newblock {\em IEEE Computer Graphics and Applications}, 38(6):125--133, 2018.

\bibitem{Hasenbein2022}
L.~Hasenbein, P.~Stark, U.~Trautwein, A.~C.~M. Queiroz, J.~Bailenson, J.-U.
  Hahn, and R.~Göllner.
\newblock Learning with simulated virtual classmates: Effects of social-related
  configurations on students’ visual attention and learning experiences in an
  immersive virtual reality classroom.
\newblock {\em Computers in Human Behavior}, 133:107282, 2022.

\bibitem{Heyse2020}
J.~Heyse, M.~Torres~Vega, T.~De~Jonge, F.~De~Backere, and F.~De~Turck.
\newblock A personalised emotion-based model for relaxation in virtual reality.
\newblock {\em Applied Sciences}, 10(17), 2020.

\bibitem{Hill2005}
V.~Hill.
\newblock Through the past darkly: a review of the british ability scales
  second edition.
\newblock {\em Child and Adolescent Mental Health}, 10(2):87--98, 2005.

\bibitem{Jackson2004}
S.~A. Jackson and R.~Eklund.
\newblock {\em The Flow Scales Manual}.
\newblock Publishers Graphics, 2004.

\bibitem{Kim2016}
J.~Kim, C.~Candido, L.~Thomas, and R.~{de Dear}.
\newblock Desk ownership in the workplace: The effect of non-territorial
  working on employee workplace satisfaction, perceived productivity and
  health.
\newblock {\em Building and Environment}, 103:203--214, 2016.

\bibitem{Latini2021}
A.~Latini, E.~{Di Giuseppe}, M.~D'Orazio, and C.~{Di Perna}.
\newblock Exploring the use of immersive virtual reality to assess occupants’
  productivity and comfort in workplaces: An experimental study on the role of
  walls colour.
\newblock {\em Energy and Buildings}, 253:111508, 2021.

\bibitem{Lee2019}
H.~Lee, S.~Je, R.~Kim, H.~Verma, H.~Alavi, and A.~Bianchi.
\newblock Partitioning open-plan workspaces via augmented reality.
\newblock {\em Personal and Ubiquitous Computing}, pp. 1--16, 2019.

\bibitem{Li2022}
J.~Li, L.~Woik, and A.~Butz.
\newblock Designing mobile mr workspaces: Effects of reality degree and spatial
  configuration during passenger productivity in hmds.
\newblock {\em Proc. ACM Hum.-Comput. Interact.}, 6(MHCI), sep 2022.

\bibitem{Medeiros2022}
D.~Medeiros, M.~McGill, A.~Ng, R.~McDermid, N.~Pantidi, J.~Williamson, and
  S.~Brewster.
\newblock From shielding to avoidance: Passenger augmented reality and the
  layout of virtual displays for productivity in shared transit.
\newblock {\em IEEE Transactions on Visualization and Computer Graphics},
  28(11):3640--3650, 2022.

\bibitem{Mostajeran2023}
F.~Mostajeran, F.~Steinicke, S.~Reinhart, W.~Stuerzlinger, B.~E. Riecke, and
  S.~K{\"u}hn.
\newblock Adding virtual plants leads to higher cognitive performance and
  psychological well-being in virtual reality.
\newblock {\em Scientific Reports}, 13(1):8053, 2023.

\bibitem{Raven2003}
J.~Raven and J.~Raven.
\newblock {\em Raven Progressive Matrices}, pp. 223--237.
\newblock Springer US, Boston, MA, 2003.

\bibitem{Ruvimova2020}
A.~Ruvimova, J.~Kim, T.~Fritz, M.~Hancock, and D.~C. Shepherd.
\newblock "transport me away": Fostering flow in open offices through virtual
  reality.
\newblock In {\em Proceedings of the 2020 CHI Conference on Human Factors in
  Computing Systems}, CHI '20, p. 1–14. Association for Computing Machinery,
  New York, NY, USA, 2020.

\bibitem{Schaffer2013}
O.~Schaffer.
\newblock Crafting fun user experiences: A method to facilitate flow.
\newblock {\em Human Factors International}, 2013.

\bibitem{Schmuecker2021}
V.~Schmücker, T.~J. Eiler, A.~Grünewald, S.~Forstmeier, F.~Grensing,
  C.~Gießer, and R.~Brück.
\newblock Customizable memory training in virtual reality with personal
  memoirs.
\newblock In {\em 2021 IEEE International Conference on Artificial Intelligence
  and Virtual Reality (AIVR)}, pp. 189--193, 2021.

\end{thebibliography}

\appendix

\section{Interview Questions}
\label{app_interview_questions}

\begin{enumerate}
    \item What is your preferred work environment? Why?
    \item What aspects of this environment help you stay focused and be productive?
    \item Do you make any changes or adjustments to your work environment before you start working in order to focus better? If yes, what are they?
    \item Would you say that your work environment preferences change from time to time? If yes, how?
    \item What do you find distracting in a work environment? Why?
    \item Imagine you are given an advanced, perfectly comfortable VR headset where the virtual environment feels incredibly realistic. So now you have an unlimited choice of possible work environments. Where would you choose to work? What is your ideal work environment? Why?
    \item What aspects of the work environment help you to be calm and relaxed? Why?
    \item  How do the following aspects influence your performance:

   a. The type of environment (urban, natural, etc.)
   
   b. Ambient sounds
   
   c. Music
   
   d. Presence of familiar or personal objects
   
   e. Lighting (intensity and colour)
   
   f. Presence of nearby clocks, timers in applications and software, etc.
\end{enumerate}

\section{Semi-structured Interview}
\label{app_end_interview}

\begin{enumerate}
    \item  What is your opinion about each of the environments?

   a. Dark Room
   
   b. Empty Room
   
   c. Furnished Room
   \item Did you feel immersed in the environments?
   \item In which environment did you find it the easiest to focus? Why?
   \item In which environment did you have the most trouble focusing? Why?
   \item What did you think about the matrix pattern questions? Did you find them difficult to solve?
   \item What do you think about your performance? Do you think you did well?
   \item What aspects of the environments do you consider could have influenced your performance in the matrix pattern questions? Why?
   \item  Which of the environments from the ones you have experienced would you consider to be the best for working? Why?
   \item Do you have any additional feedback about the experiment?
\end{enumerate}

\end{document}